\documentclass[aps,twocolumn,pre,showpacs,balancelastpage,amssymb,groupedaddress]{revtex4}
\usepackage{graphicx}
\usepackage{amsmath}

\begin{document}
\title{Lyapunov Generation of Entanglement and the Correspondence Principle}
\author{C.~Petitjean$^1$ and Ph.~Jacquod$^{1,2}$}
\affiliation{$^1$ D\'epartement de Physique Th\'eorique,
Universit\'e de Gen\`eve, CH-1211 Gen\`eve 4, Switzerland \\
$^2$ Department of Physics, University of Arizona, 1118 E. Fourth Street, 
Tucson, AZ 85721}
\begin{abstract}
We show how a classically vanishing interaction generates
entanglement between two initially nonentangled particles,
without affecting their classical dynamics. 
For chaotic dynamics, the rate of entanglement 
is shown to saturate at the Lyapunov exponent of the 
classical dynamics as the interaction strength increases.
In the saturation regime, 
the one-particle Wigner function follows classical dynamics
better and better as one goes deeper and deeper in the semiclassical
limit. This demonstrates
that quantum-classical correspondence at the microscopic level requires neither
high temperatures, nor coupling to a large number of external degrees of 
freedom.
\end{abstract}

\pacs{05.45.Mt,03.65.Ud,05.70.Ln,03.67.-a}

\maketitle

In the decades since its inception, no observed phenomenon, nor experimental
result ever contradicted quantum theory. Yet, the world surrounding us, 
though being made out of quantum mechanical building blocks, behaves 
classically most of the time. This suggests that, one way or another,
classical physics emerges out of quantum mechanics. Today's common
understanding of this quantum-classical
correspondence is based on the realization that no
finite-sized system is ever fully
isolated. It is then hoped that a large regime of parameters exists where the 
coupling of the system to external degrees of freedom (to be called the environment from now on)
destroys quantum interferences without modifying the system's classical 
dynamics.
Indeed, such a coupling usually induces loss
of coherence on a time scale much shorter than it relaxes the 
system \cite{Joos03,Zurek03}.

The standard approach to decoherence starts from
a master equation valid in the regime of 
weak system-environment coupling \cite{Joos03,Zurek03}. 
The master equation determines the time-evolution of the system's Wigner
function $W({\bf p},{\bf q}) = (2 \pi \hbar)^{-d} \int {\rm d} {\bf x} \exp[i {\bf p} {\bf x}]
\rho({\bf q}+{\bf x}/2,{\bf q}-{\bf x}/2)$ ($\rho$ is the system's density matrix)
as
\begin{eqnarray}\label{Wigner_o}
\partial_t W &=& \Big\{ H,W \Big\} + \sum_{n \ge 1}
\frac{(i\hbar)^{2n}}{2^{2n} (2n+1)!} \partial_q^{2n+1} V \partial_p^{2n+1} W
 \nonumber \\
&& + 2 \gamma \partial_p (pW) + D \partial^2_{p} W.
\end{eqnarray}
The first term on the right-hand side of Eq.~(\ref{Wigner_o}) is the classical Poisson
bracket. The second term, written here for the case of a momentum-independent
potential $V({\bf q})$, 
exists already in closed systems and generates quantum corrections
to the dynamical evolution of $W$. This term starts to become comparable to the Poisson
bracket at the Ehrenfest
time $\tau_{\rm E} = \lambda^{-1} \ln N$, where $\lambda$ is the Lyapunov exponent of the
classical dynamics, and
$N$ the size of the system's Hilbert space. The two terms on the second line of Eq.~(\ref{Wigner_o})
are induced by the coupling to the environment. In the limit of weak coupling, $\gamma \rightarrow 0$,
but finite diffusion constant, $D \propto \gamma T = {\rm Cst}$, the friction term vanishes,
leaving the classical dynamics unaffected. This requires to consider the high
temperature limit. Simultaneously, for large enough $D$, 
the momentum diffusion term 
induces enough noise so as to kill the quantum corrections before they become important.
The time-evolution of $W$ is then solely governed by the classical Poisson bracket,
that is to say, classical dynamics emerges out of quantum mechanics.
Refs.~\cite{Habib,Davidovich} provided for a numerical illustration of this scenario.

Our motivations in this article are as follows.
First, it is unclear how generic the above scenario is, since it is based on a
master equation derived under restrictive assumptions, for instance
on the environment, the dimensionality of the system 
or the strength of the coupling between system and environment
\cite{Joos03,Zurek03}. Also, it formally requires
to consider infinite temperatures. Moreover, and with the
specific exception of the kicked harmonic oscillator investigated in 
Ref.~\cite{Davidovich},
there is not much analytical understanding of the decoherence process in
generic dynamical systems, i.e. except for the regular case, 
master equations are usually integrated numerically. Second,
claims have been made of an entropy production due to
the coupling to environmental degrees of freedom governed by the system's
exponent $\lambda$ \cite{Zurek03,entropy}, without clear analytical 
derivation, nor strong numerical evidence \cite{caveat1}. A Lyapunov decay of 
the fidelity has recently been analytically predicted \cite{Jal01} and 
numerically observed \cite{Jac01}, however, decoherence and fidelity are 
two different things, especially in the generic situation where the system 
and environment Hamiltonians do not commute with the coupling 
Hamiltonian \cite{Jac04,Cuc03}. 

We revisit these issues and consider two interacting 
quantized dynamical systems. Entanglement generation between two 
particles has already
been considered in Refs.~\cite{furuya98,miller99,tanaka02,Jac04,Prosen05}.
All results to date are consistent with the scenario proposed in Ref.~\cite{Jac04},
according to which bipartite entanglement results from two contributions: (i) a quantum-mechanical one,
which depends on the coupling strength between the two systems, 
and (ii) a dynamical
one, which, in chaotic systems, is determined by the total system's 
spectrum of Lyapunov exponents. 
The entanglement generation rate is given by the weakest of the coupling 
strength and the Lyapunov
exponent. It has to be pointed out that this picture holds in the regime of classically weak but
quantum-mechanically strong coupling (this will be made quantitative below). 
For regular systems, entanglement generation is slower than for chaotic ones,
typically power-law in time \cite{Jac04,Prosen05}.

The purpose of this letter is threefold. First, we address the problem of
decoherence and bipartite 
entanglement from a microscopic point of view, i.e. without relying on 
effective differential equations.
This allows for a clear identification of the regime of validity of our theory.
Second, we give strong numerical evidences for
the existence of a Lyapunov regime of entanglement in bipartite 
systems (the numerical
evidences presented in Ref.~\cite{miller99} were challenged in 
Ref.~\cite{tanaka02}). 
Third, we discuss our results from the point of view of
the quantum-classical correspondence, and present numerical phase-space 
dynamics 
results showing that this correspondence is fully achieved in the regime
of Lyapunov entanglement. This is, we believe, 
the first clear microscopic 
illustration of the quantum-classical correspondence in a generic 
chaotic system. 

As our starting point, we consider the Hamiltonian
\begin{equation}\label{2hamiltonian}
{\cal H} = H_1 \otimes I_2 + I_1 \otimes H_2 + \hbar \; {\cal U}.
\end{equation}
We take chaotic one-particle Hamiltonians $H_{1,2}$. 
We specify that the interaction potential ${\cal U}$ is smooth, varying over a distance much 
larger than the particles' de Broglie wavelength $\sigma$, and that it depends only on the
distance between the particles. Planck's constant 
in front of ${\cal U}$ in Eq.~(\ref{2hamiltonian})
makes it explicit that we consider a semiclassically vanishing two-particle interaction, i.e.
the classical Hamiltonian corresponding to Eq.~(\ref{2hamiltonian}) does not
couple the two particles. 
Our goal is to calculate the purity ${\cal P}(t)={\rm Tr}_1[\rho_1^2(t)]$ of the reduced density matrix
$\rho_1(t)={\rm Tr}_2 \rho(t)$. In the situation we consider of a unitary two-particle dynamics acting on an initially pure two-particle state, ${\cal P}(t)$
is a good measure of entanglement, which varies between 1 for product states and 0 for maximally entangled states. The calculation proceeds along the lines of Ref.~\cite{Jac04},
and here we only sketch it. A similar semiclassical approach has been applied
to a stochastic Schr\"odinger equation in Ref.~\cite{Kolovsky}.

In the initial two-particle product state we take, each particle is in a
Gaussian wavepacket $\psi_{1,2}({\bf y}) = (\pi \sigma^2)^{-d/4} \exp[i {\bf
    p}_{1,2} \cdot ({\bf y}-{\bf r}_{1,2})/\hbar-|{\bf y}-{\bf r}_{1,2}|^2/2 
\sigma^2]$. The two-particle density matrix evolves according to $\rho(t) = 
\exp[-i {\cal H} t/\hbar] |\psi_1; \psi_2 \rangle
\langle \psi_1; \psi_2| \exp[i {\cal H} t/\hbar]$
This time-evolution is evaluated semiclassically by means of the 
semiclassical two-particle propagator
\vskip -5mm
\begin{equation}\label{2prop}
\langle {\bf x}_1, {\bf x}_2| e^{-i{\cal H}t/\hbar} |{\bf y}_1,
{\bf y}_2 \rangle =  \sum_{s,s'} C_{s,s'}^{1/2} 
e^{i [(S_s +
S_{s'})/\hbar + {\cal S}_{s,s'}]} 
\end{equation}
which is expressed as a sum over pairs of classical trajectories, labelled $s$
and $s'$, respectively 
connecting ${\bf y}_1$ to ${\bf x}_1$ and ${\bf y}_2$ to ${\bf x}_2$
in the time $t$. Because of our assumption of a semiclassically vanishing 
coupling, these classical trajectories are determined by the 
one-particle Hamiltonians.
Each pair of such trajectories gives a contribution   
weighted by $C_{s,s'}$, the inverse of the determinant of the 
stability matrix on $s$ and $s'$, and oscillating with 
one-particle (denoted by $S_s$ and $S_{s'}$) 
and two-particle (denoted by ${\cal S}_{s,s'}= 
\int_0^t dt' {\cal U}[{\bf q}^{(1)}_s (t'), {\bf q}^{(2)}_{s'}(t')]$) 
action integrals accumulated by the first and second particles
along $s$ and $s'$ respectively.
In the regime of semiclassically vanishing coupling we consider, 
the one-particle actions generate much faster oscillations
than their two-particle counterpart. Accordingly,
our approach relies on stationary phase
conditions imposed on the one-particle actions.
In Eq.~(\ref{2prop}),
Maslov indices have been omitted since they drop
out of the calculation.

To leading order in $\hbar_{\rm eff}=2 \pi/N_{1,2}$ ($N_i$ is the size of the $i^{\rm th}$ system's
Hilbert space), our semiclassical calculation gives 
the time-evolution of the purity as 
\begin{eqnarray}\label{purity}
{\cal P}(t) &\simeq& \sum_{i=1,2} \alpha_i \Theta(t>\tau_i)
\exp[-\lambda_i t]  + \exp[-2 \Gamma t]  \nonumber \\
&& + \Theta(t>\tau^{(1)}_{\rm E}) N_1^{-1}
+ \Theta(t>\tau^{(2)}_{\rm E}) N_2^{-1}.
\end{eqnarray}
The first, classical term decays with the Lyapunov
exponents $\lambda_{1,2}$ \cite{caveat2}. It does not exist
at short times, $t<\tau_i = \lambda_i^{-1} \ln[\lambda_i/\sigma^2 G_i]$, 
($G_i = \int {\rm d}t' \langle \partial_q^{(i)} {\cal U}[{\bf q}^{(1)}_s(0), 
{\bf q}^{(2)}_{s'}(0)]\; \partial_q^{(i)} {\cal U}[{\bf q}^{(1)}_s(t'),
{\bf q}^{(2)}_{s'}(t')] \rangle$), and has prefactors $\alpha_i = {\cal O}(1)$.
The second term is the standard, interaction-dependent
quantum term with $\Gamma = \int_0^t {\rm d}t'
\langle {\cal U}[{\bf q}^{(1)}_s(0), {\bf q}^{(2)}_{s'}(0)]\;
{\cal U}[{\bf q}^{(1)}_s(t'),{\bf q}^{(2)}_{s'}(t')] \rangle$,
assuming a fast decay of correlations. Being given by the correlator
of a classical potential evaluated along classical trajectories,
$\Gamma$ does not depend on $\hbar$. Finally, the third and fourth, 
saturation terms in Eq.~(\ref{purity}) set in after
$\tau_{\rm E}^{(i)} = \lambda^{-1}_i \ln N_i$.

The validity of our approach is given by
$\delta_2 \le \Gamma \le B_2$, where $B_2$ and $\delta_2=B_2/(N_1 N_2)$ 
are the two-particle bandwidth and level spacing respectively
\cite{Jac01}. In this range, 
${\cal U}$ is quantum-mechanically strong
as individual levels are broadened beyond their average spacing, 
but classically weak, as $B_2$ is unaffected by ${\cal U}$ \cite{Jal01,Jac01}. 
We note that our semiclassical approach preserves the properties of the 
density matrix ${\rm Tr}_1[\rho_1(t)]=1$, $\rho_1=\rho_1^\dagger$, 
as well as the symmetry ${\rm Tr}_1[\rho_1^2(t)]={\rm Tr}_2[\rho_2^2(t)]$.

Eq.~(\ref{purity}) expresses the decay of ${\cal P}(t)$
as a sum over dynamical, purely classical contributions, and quantal ones, 
depending on the 
interaction strength. Because the decaying terms are exponential
and have prefactors of order unity, one has for $t>\tau^{(1,2)}_{\rm E},\tau_{1,2}$
\vskip -5mm
\begin{equation}\label{puritysum}
{\cal P}(t) \simeq
\exp[-{\rm min} (\lambda_1,\lambda_2,2 \Gamma ) t] + N_1^{-1} + N_2^{-1}.
\end{equation}
Eq.~(\ref{puritysum}) reconciles the results of Refs.~\cite{miller99} and \cite{tanaka02}. 
Its regime of validity, $\delta_2=B_2/(N_1 N_2) \le 
\Gamma \le B_2$, is parametrically large in the semiclassical limit $N_{1,2} 
\rightarrow \infty$. The same approach also applies to regular systems, in
which case the exponentially decaying Lyapunov terms are replaced by 
power-law decaying terms \cite{Jac04,Prosen05}.

We now discuss the connection of our main result, Eq.~(\ref{puritysum}),
to Eq.~(\ref{Wigner_o}). The purity measures the weight of off-diagonal
elements of $\rho_1(t)$, and hence of the importance of coherent
effects. In the regime $2 \Gamma \gg \lambda_1=\lambda_2$, 
${\cal P}(t)$ reaches its minimal value at the Ehrenfest time.
Thus, quantum effects [the second term
on the right-hand side of Eq.~(\ref{Wigner_o})] are killed
before they have a chance to appear. In that regime, one therefore expects the
quantum-classical correspondence to become complete in the semiclassical
limit $N_{1,2} \rightarrow \infty$. We will show below numerical evidences
supporting this reasoning.

\begin{figure}
\includegraphics[width=7.8cm]{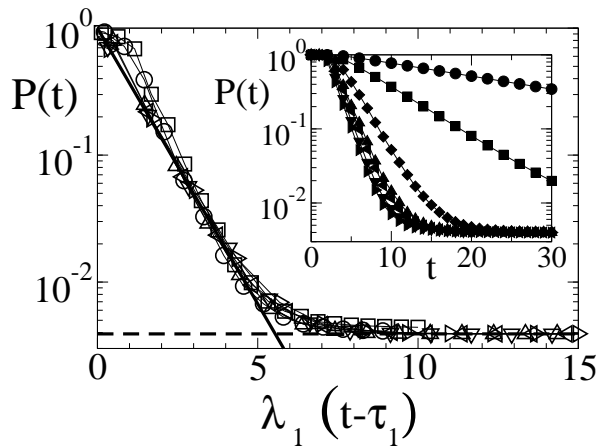}
\caption{\label{fig1}
Main plot: Purity of the reduced 
density matrix for $N_1=N_2=512$, $K_1=K_2 \in[4,12]$, and
$\epsilon=4$ giving $2 \Gamma = 13.6 \gg \lambda_1 = \lambda_2$. Data have been calculated from
20 different initial states. The time axis has been shifted by the onset
time $\tau_1$ (see text) and rescaled with $\lambda_1 \in [0.5,1.35]$.
The full line indicates $ \propto \exp[-\lambda_1 t]$, and
the dashed line gives the asymptotic saturation ${\cal P}(\infty) = 2N_1^{-1}$.
Inset: Purity for $K_1=K_2=5.09$ for
$ \epsilon = 0.2$ (circles), $0.4$ (squares), $0.8$ (diamonds), 
 $1.6, 2,3$ and $4$ (triangles).\\[-6mm]}
\end{figure}

To numerically check our results, we consider Hamiltonian 
(\ref{2hamiltonian}) for two coupled kicked rotators \cite{Izr90}
\begin{subequations}
\label{krot}
\begin{eqnarray}
H_i & = & p_i^2 / 2 + K_i \cos(x_i) \; \sum_n \delta(t-nT),\\
{\cal U} & = & \epsilon \; \sin(x_1-x_2-0.33) \; \sum_n \delta(t-nT).
\end{eqnarray}
\end{subequations}
The interaction potential ${\cal U}$ is long-ranged with a 
strength $\epsilon$ and acts at the same time as the kicks.
Upon increasing $K_i$ the classical dynamics of the $i^{\rm th}$
particle varies from fully
integrable ($K_i=0$) to fully chaotic [$K_i\agt 7$, with Lyapunov exponent
$\lambda_i\approx\ln (K_i/2)$]. For $1<K_i<7$ the dynamics is mixed. We will vary 
$K_{1,2} \in [3,12]$ to get a maximal variation of $\lambda_i$, while
making sure that both $\psi_1$ and $\psi_2$ lie in the chaotic sea.
We follow the usual quantization procedure on the torus $x,p\in(-\pi,\pi)$. 
The bandwidth and level spacing are given by
$B_2 = 4 \pi$, $\delta_2 = 4 \pi/N^2$, and we numerically extracted
$\Gamma \simeq 0.43\epsilon^2$ from exact diagonalization calculations
of the local spectral density of states. The time evolved density matrix 
is computed by means of fast fourier transforms \cite{Izr90}. 
The algorithm requires only ${\cal O}(N \ln N)$ operations, which
allowed us to reach system sizes up to $N_{1,2}=2048$, more than one order
of magnitude larger than any previously investigated case. 

The behavior of ${\cal P}(t)$ is shown in Fig.~\ref{fig1}. First, it is seen
that as $\epsilon$ increases, the rate of entanglement generation also 
increases, up to some value $\epsilon_c$, after which it saturates. We have 
found that (i) prior to saturation, ${\cal P}(t)$ decays exponentially with a 
rate $\approx 0.85 \epsilon^2$, provided $\Gamma = 0.43 \epsilon^2 > 
\delta_2 = 4 \pi/N^2$ is satisfied, and that (ii)
$\epsilon_c$ behaves consistently with Eq.~(\ref{puritysum}). Second,
Fig.~1 shows how ${\cal P}(t)$ behaves for
fixed $\epsilon > \epsilon_c$ upon variation of the Lyapunov exponents
$\lambda_1 = \lambda_2$. The rescaling of the time axis $t \rightarrow
\lambda_1 t$ allows to bring together six curves with 
$\lambda_1 \in [0.5,1.35]$, varying by almost a factor three. Third, Fig.~1
shows that in the chaotic regime considered here with $N_1 = N_2$, 
${\cal P}(t \rightarrow \infty) = 2 N_1^{-1}$. 
These numerical data fully confirm our main results, Eqs.~(\ref{purity}) and (\ref{puritysum}).

\begin{figure*}
\includegraphics[width=7.6cm]{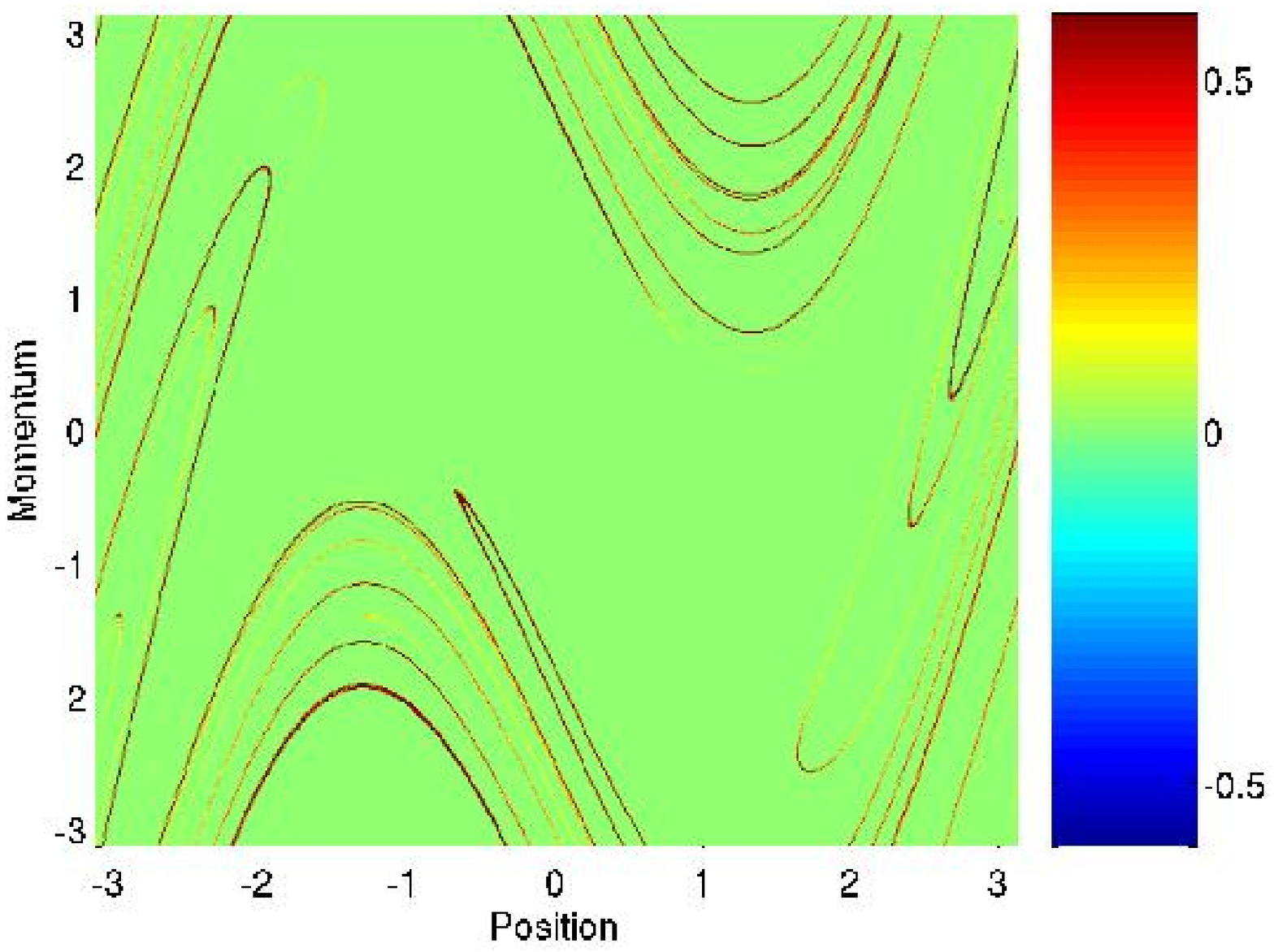}
\includegraphics[width=7.6cm]{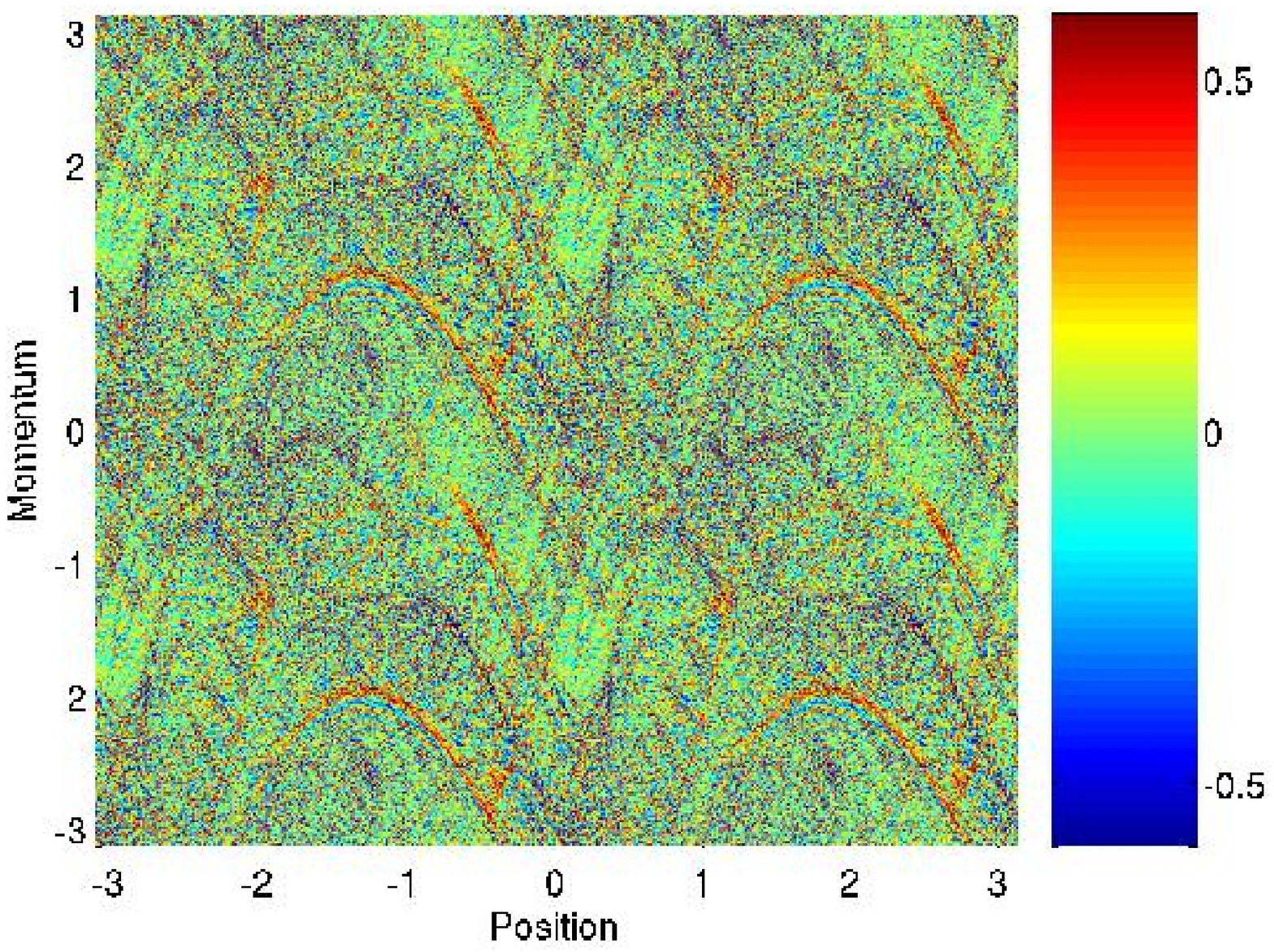} \\
\includegraphics[width=7.6cm]{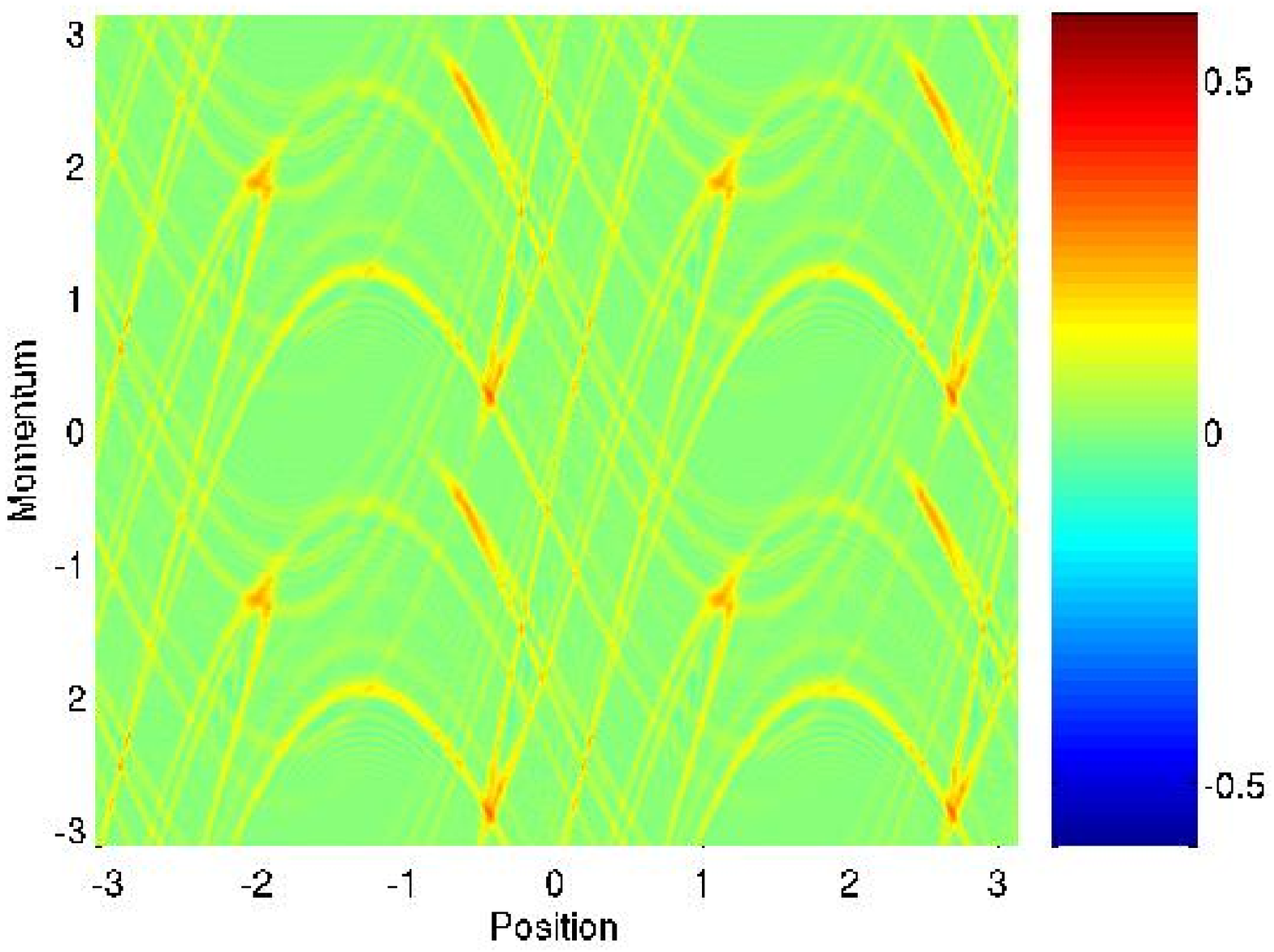}
\includegraphics[width=7.6cm]{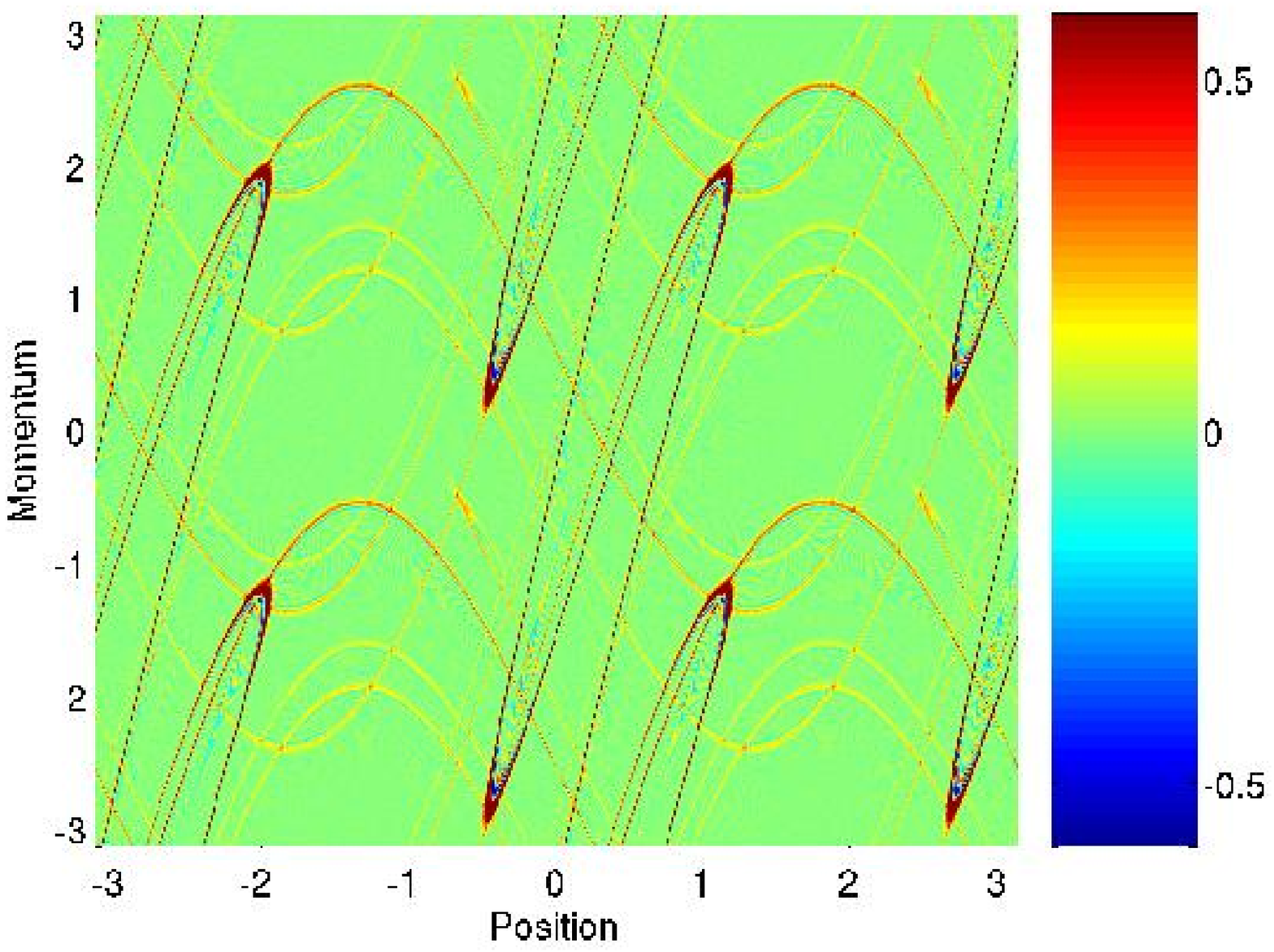}
\caption{\label{fig2} (Color online) Phase-space plots for a classical distribution (top left),
uncoupled (top right) and coupled (bottom left and right, $\epsilon=4$) quantum
Wigner distributions, after five iterations of the kicked 
rotator map of Eqs.~(\ref{2hamiltonian}) and (\ref{krot}). In all cases, the system has $K_1=3.09$, and
the initial distributions are Gaussian centered in the chaotic sea 
at $(x,p)=(1,2)$. Bottom panels: Wigner functions for the 
quantum system coupled to a second kicked rotator
with $K_2 = 100$. One has $2 \Gamma=13.6 > \lambda_2 \gg \lambda_1$, 
so that the purity behaves as ${\cal P}(t) \simeq \exp[-\lambda_1 t]$.
The left panel has $N_1=N_2=512$ and the right panel has
$N_1=N_2=2048$. The presence of ghost images in the
Wigner function is an artifact of the boundary conditions 
\cite{Dittrich}.\\[-8mm]}
\end{figure*}

We next turn our attention to the quantum-classical correspondence in
phase space. We compare in
Fig.~\ref{fig2} the Liouville evolution of a classical distribution
with that of the Wigner function $W_1({\bf p},{\bf q};t) = 
(2 \pi \hbar)^{-d} \int {\rm d} {\bf x} \exp[i {\bf p} {\bf x}]
\rho_1({\bf q}+{\bf x}/2,{\bf q}-{\bf x}/2;t)$. The latter
is quantum-mechanically evolved 
from a localized wavepacket with the same initial location and extension as
the classical distribution. Three quantum phase-space plots are
shown: (i) (top right) for a free system, $\epsilon=0$; (ii) and (iii) 
(bottom left and right) for a coupled system
$\epsilon = 4$, in the regime ${\cal P}(t) \simeq \exp[-\lambda_1 t]$. The bottom left panel
has a system size $N_1=N_2=512$ while the bottom right panel has $N_1=N_2=2048$. All plots
show phase-space distributions after 5 kicks, a duration comparable 
to $\tau_{\rm E}$.
Two things are clear from these figures. First, a coupling is necessary and sufficient
to achieve phase-space quantum-classical correspondence. Second, the
correspondence becomes better as we move deeper in the semiclassical
regime, {\it even though the interaction Hamiltonian vanishes in that limit !}

One key issue is whether the observed classical entanglement rate 
translates into a Lyapunov decoherence rate for systems
coupled to a true environment. The latter differs from a
coupling to a single particle in that it has much shorter 
time scales, it has a much bigger Hilbert space, and
it cannot be initially prepared in a pure Gaussian wavepacket.
We can take these conditions into account in our semiclassical approach 
by considering (i) $\lambda_2 \gg \lambda_1$, (ii) $N_2 \rightarrow \infty$
and (iii) taking an initial mixed environment density matrix 
$\rho_{\rm env}=\sum_a |\phi_a|^2 |a \rangle \langle a |$, with
$\langle {\bf x} |a\rangle$ being nonoverlapping 
Gaussian wavepackets. The result is that Eq.~(\ref{purity}) is replaced by
\begin{eqnarray}\label{decoherence}
{\cal P}(t) &\simeq& \alpha_1 \Theta(t>\tau_1)
\exp[-\lambda_1 t] + \exp[-2 \Gamma t] \nonumber \\
&& + \Theta(t > \tau^{(1)}_{\rm E}) N_1^{-1}.
\end{eqnarray}
The dynamical Lyapunov decay of the purity seems to survive in the case
of a particle coupled to an environment. We have obtained numerical
confirmation of Eq.~(\ref{decoherence}) which we do not present here.

We stress in conclusion
that one advantage of our approach is that ${\cal P}(t)$ is directly 
calculated,
without the step of numerically integrating a 
differential equation for $\rho_1(t)$.
Future works should focus on multipartite entanglement and decoherence by 
an environment consisting of many coupled dynamical systems.

This work has been supported by the Swiss National Science Foundation.


\begin{thebibliography}{99}
\bibitem{Joos03} E. Joos, H.D. Zeh, C. Kiefer, D. Giulini, J. Kupsch, 
I.-O. Stamatescu, {\it Decoherence and the Appearance of a Classical World in 
Quantum Theory}, 2nd Ed. (Springer, Berlin 2003).
\bibitem{Zurek03} W.H. Zurek, Rev. Mod. Phys. {\bf 75}, 715 (2003).
\bibitem{Habib} S. Habib, K. Shizume,
and W.H. Zurek, Phys. Rev. Lett. {\bf 80}, 4361 (1998).
\bibitem{Davidovich} F. Toscano, R.L. de Matos Filho, and
L. Davidovich, Phys. Rev. A {\bf 71}, 010101(R) (2005).
\bibitem{entropy} A.K. Pattanayak, Phys. Rev. Lett. {\bf 83}, 4526 (1999);
  D. Monteoliva and J.P. Paz, Phys. Rev. Lett. {\bf 85}, 3373 (2000).
\bibitem{caveat1} Refs.~\cite{entropy} show entropy production rates 
without varying the Lyapunov exponent.
\bibitem{Jal01} R.A. Jalabert and H.M. Pastawski, Phys. Rev. Lett. 
{\bf 86}, 2490 (2001).
\bibitem{Jac01} Ph. Jacquod, P.G. Silvestrov, and C.W.J. Beenakker,
Phys. Rev. E {\bf 64}, 055203(R) (2001).
\bibitem{Cuc03} F.M. Cucchietti, D.A.R. Dalvit, J.P. Paz, and W.H. Zurek,
  Phys. Rev. Lett. {\bf 91}, 210403 (2003).
\bibitem{Jac04} Ph. Jacquod, Phys. Rev.Lett. {\bf 92}, 150403 (2004).
\bibitem{furuya98} K. Furuya, M.C. Nemes, and G.Q. Pellegrino, Phys. Rev. 
Lett. {\bf 80}, 5524 (1998); J.N. Bandyopadhyay and A. Lakshminarayan, Phys. Rev. Lett {\bf 89}, 060402 
(2002); M. \v Znidari\v c and T. Prosen, J. Phys. A {\bf 36}, 
2463 (2003); R.M. Angelo and K. Furuya, Phys. Rev. A {\bf 71}, 042321 (2005); 
M. Lombardi and A. Matzkin, quant-ph/0506188.
\bibitem{miller99} P.A. Miller and S. Sarkar, Phys. Rev. E {\bf 60}, 1542 
(1999).
\bibitem{tanaka02} A. Tanaka, H. Fujisaki, and T. Miyadera, Phys. Rev. E 
{\bf 66}, 045201(R) (2002).
\bibitem{Prosen05} M. \v Znidari\v c and T. Prosen, 
Phys. Rev. A {\bf 71}, 032103 (2005).
\bibitem{Kolovsky} A.R. Kolovsky, Phys. Rev. Lett. {\bf 76}, 340 (1996).
\bibitem{caveat2} There are subtleties related to averaging, so that
$\lambda_{1,2}$ are somewhat smaller than, but proportional to the
Lyapunov exponents, see P.G. Silvestrov, J. Tworzydlo, and C.W.J. Beenakker,
Phys. Rev. E {\bf 67}, 025204(R) (2003).
\bibitem{Izr90} F.M. Izrailev, Phys. Rep. {\bf  196}, 299 (1990).
\bibitem{Dittrich} A. Arguelles and T. Dittrich, Physica A {\bf 356}, 
72 (2005).

\end{thebibliography}
\end{document}